\documentclass[aps,prl,showpacs,twocolumn,superscriptaddress]{revtex4}
\usepackage{bm,color}
\usepackage{graphicx}
\usepackage{amsmath}

\begin{document}
\title {Spin-Wave Modes and Their Intense Excitation Effects in Skyrmion Crystals}

\author{Masahito Mochizuki}
\affiliation{Department of Applied Physics, University of Tokyo,
Tokyo 113-8656, Japan}
\affiliation{Multiferroics Project, ERATO, Japan Science and Technology
Agency (JST), Tokyo 113-8656, Japan}


\begin{abstract}
We theoretically study spin-wave modes and their intense excitations activated by microwave magnetic fields in the skyrmion-crystal phase of insulating magnets by numerically analyzing a two-dimensional spin model using the Landau-Lifshitz-Gilbert equation. Two peaks of spin-wave resonances with frequencies of $\sim$1 GHz are found for in-plane a.c. magnetic field where distribution of the out-of-plane spin components circulates around each skyrmion core. Directions of the circulations are opposite between these two modes, and hence the spectra exhibit salient dependence on the circular polarization of irradiating microwave. A breathing-type mode is also found for out-of-plane a.c. magnetic field. By intensively exciting these collective modes, melting of the skyrmion crystal accompanied by a red shift of the resonant frequency is achieved within nano seconds.
\end{abstract}
\maketitle
Competing interactions in magnets often cause nontrivial spin textures such as ferromagnetic domains and magnetic bubbles, which have attracted a great deal of interest from the viewpoints of both fundamental science and technical applications in the field of spintronics~\cite{Malozemoff79,BaderRMP06}. In particular, response dynamics of such magnetic structures under external fields is an issue of vital importance because its understanding is crucial for their manipulations.

Skyrmion, a nontrivial swirling spin structure carrying a topological quantum number, is one of the interesting examples of such spin textures. It was originally proposed by Skyrme to account for baryons in nuclear physics in 1960's as a quasiparticle excitation with spins pointing in all directions to wrap a sphere~\cite{Skyrme61,Skyrme62}, and was recently realized experimentally in two-dimensional condensed matter systems, e.g., quantum Hall ferromagnets~\cite{Sondhi93,Abolfath97}, ferromagnetic monolayers~\cite{Heinze11}, and doped layered antiferromagnets~\cite{Raicevic11}.

Formation of skyrmion crystal (SkX) was theoretically predicted in Dzyaloshinskii-Moriya (DM) ferromagnets without inversion symmetry~\cite{Bogdanov89,Bogdanov94}, and was indeed observed in the $A$-phase of metallic chiral magnets MnSi~\cite{Muhlbauer09,Pfleiderer10} and Fe$_{1-x}$Co$_x$Si~\cite{Munzer10} by neutron-scattering experiments as a triangular lattice of skyrmions with spins antiparallel to the applied magnetic field at the skyrmion centers and parallel at their peripheries. A recent Monte-Carlo study found a greater stability of the SkX phase in thin films~\cite{YiSD09}. This prediction was confirmed by the real-space observation of skyrmion triangular lattice in Fe$_{0.5}$Co$_{0.5}$Si thin films using the Lorentz force microscopy in a wide temperature and magnetic-field range~\cite{YuXZ10N}.

Typically the skyrmion is 10-100 nm in size, which is determined by the ratio of DM interaction and exchange coupling and is much smaller than magnetic bubbles. Moreover, recent experiments found that the skyrmion is stable even near/above the room temperature~\cite{YuXZ10M}, and can be manipulated by much lower electric currents than ferromagnetic domain walls~\cite{Jonietz10,Everschor11}. These properties, i.e., small size, high operational temperature, and low threshold field, are advantageous for technical application to high-density data storage devices. Therefore, understanding of the dynamics of skyrmions and SkX under external fields is an important issue~\cite{SKHall}.

In this Letter, we theoretically study collective spin dynamics in the SkX phase of insulating ferromagnets with DM interaction by numerical simulations of the Landau-Lifshitz-Gilbert (LLG) equation under time-dependent a.c. magnetic fields. We find a couple of spin-wave resonances with frequencies $\sim$ 1 GHz for in-plane a.c. magnetic field where the out-of-plane spin components rotate around each skyrmion core. The directions of these rotations are opposite between the higher-lying and lower-lying modes, and their spectra show strong circular-polarization dependence. A breathing-type mode is also found for out-of-plane a.c. magnetic field.  Furthermore, we study intense excitation effects of these collective modes, and find a red shift of the resonant frequency and melting of the SkX within nano seconds. These findings will lead to a fast manipulation of skyrmions in nano scale using spin-wave resonances.

We start with a classical Heisenberg model on a two-dimensional square lattice~\cite{YiSD09}, which contains nearest-neighbor ferromagnetic exchange, Zeeman coupling, and DM interaction as~\cite{Bak80},
\begin{eqnarray}
\mathcal{H}&=&-J \sum_{<i,j>} \bm S_i \cdot \bm S_j 
-\left[ \bm H + \bm H'(t) \right] \cdot \sum_i \bm S_i \nonumber \\
&+&D \sum_{i} (\bm S_i \times \bm S_{i+\hat{x}} \cdot \hat{x}
+\bm S_i \times \bm S_{i+\hat{y}} \cdot \hat{y}),
\label{eqn:model}
\end{eqnarray}
where $\bm H$=(0, 0, $H_z$) is a constant external magnetic field normal to the plane, and $\bm H'(t)$ is an applied time-dependent magnetic field. The norm of the spin vector is set to be unity. We adopt $J$=1 as the energy unit and take $D$=0.09. The spin turn angle $\theta$ in the helical structure is determined by the ratio $D/J$ as $\tan \theta$=$D/(\sqrt{2}J)$, which is derived from a saddle point equation of the energy as a function of $\theta$. Our parameter set gives $\theta$=3.64$^{\circ}$ or the periodicity of $\sim$99 sites, which corresponds to the skyrmion diameter of $\sim$50 nm if we consider a typical lattice parameter of 5 $\AA$.

We study collective spin excitations of this model by numerically solving the LLG equation using the fourth-order Runge-Kutta method. The equation is given by
\begin{equation}
\frac{\partial \bm S_i}{\partial t}=
-\frac{1}{1+\alpha_{\rm G}^2} \left[
\bm S_i \times \bm H^{\rm eff}_i + \frac{\alpha_{\rm G}}{S}
\bm S_i \times (\bm S_i \times \bm H^{\rm eff}_i) \right],
\label{eq:LLGEQ}
\end{equation} 
where $\alpha_{\rm G}$ is the dimensionless Gilbert-damping coefficient. We derive a local effective field $\bm H^{\rm eff}_i$ acting on the $i$th spin $\bm S_i$ from the Hamiltonian $\mathcal{H}$ as
$\bm H^{\rm eff}_i = - \partial \mathcal{H} / \partial \bm S_i$.
All the calculations are performed for systems with $N$=288$\times$288 sites under the periodic boundary condition. We fix $\alpha_{\rm G}$=0.04 for simulations of the spectra shown in Fig.~\ref{Fig02}, while $\alpha_{\rm G}$=0.004 for others.

\begin{figure}
\includegraphics[scale=1.0]{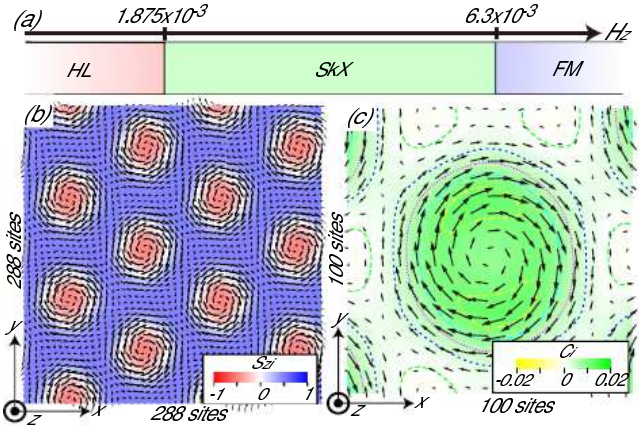}
\caption{(color). (a) Phase diagram of the Hamiltonian~(\ref{eqn:model}) at $T$=0 where HL, SkX, and FM denote helical, skyrmion-crystal, and ferromagnetic phases, respectively. (b) Spin configuration of the SkX phase with a color map of the spin $z$-axis components $S_{zi}$ at $H_z$=3.75$\times$10$^{-3}$. Spin vectors at sites ($i_x$, $i_y$) projected onto the $xy$ plane are shown by arrows for mod($i_x$, 6)=mod($i_y$, 6)=0. (c) One skyrmion is magnified with a color map of the scalar spin chiralities $C_i$.}
\label{Fig01}
\end{figure}
\begin{table}
\caption{Unit conversion table when $J$=0.4 meV.}
\begin{tabular}{c|cc}
\hline
\hline
magnetic field $H$ &  1$\times$10$^{-3}$$J$ & $\sim$3.4 mT \\
frequency $\omega$ &  0.01$J$ & $\sim$1 GHz \\
time $t$  &  1000$J^{-1}$ & $\sim$10 nsec \\
\hline
\hline
\end{tabular}
\label{tab:convtb}
\end{table}
We first study phase diagram of the model~(\ref{eqn:model}) at $T$=0 as a function of $H_z$. Starting with spin configurations obtained in the Monte-Carlo thermalization at low $T$, we further relax them by sufficient time evolution in the LLG equation, and compare their energies. As shown in Fig.~\ref{Fig01}(a), helical (HL), SkX, and ferromagnetic (FM) phases appear successively as $H_z$ increases where critical fields are $H_z$=1.875$\times$10$^{-3}$ and $H_z$=6.3$\times$10$^{-3}$, respectively. Here $H_z$=1$\times$10$^{-3}$ corresponds to $\sim$3.4 mT if we adopt a typical value of $J$=0.4 meV and $S$=1 spins (see also Table~\ref{tab:convtb}). In Fig.~\ref{Fig01}(b), we display spin configuration of the SkX phase where the in-plane components of the spin vectors at sites ($i_x$, $i_y$) are described by arrows when mod($i_x$, 6)=mod($i_y$, 6)=0. Here distribution of the spin $z$-axis components, $S_{zi}$, is shown by a color map. One skyrmion is magnified in Fig.~\ref{Fig01}(c) with a color map of the local scalar spin chiralities given by,
\begin{equation}
C_i=\bm S_i \cdot (\bm S_{i+\hat{x}} \times \bm S_{i+\hat{y}})
+\bm S_i \cdot (\bm S_{i-\hat{x}} \times \bm S_{i-\hat{y}}).
\label{eqn:SSC}
\end{equation}
The finite spin chirality is a source of the topological Hall effect~\cite{Binz08} observed in experiments~\cite{LeeM07,Neubauer09,Kanazawa11,Pfleiderer10b}.

\begin{figure}
\includegraphics[scale=1.0]{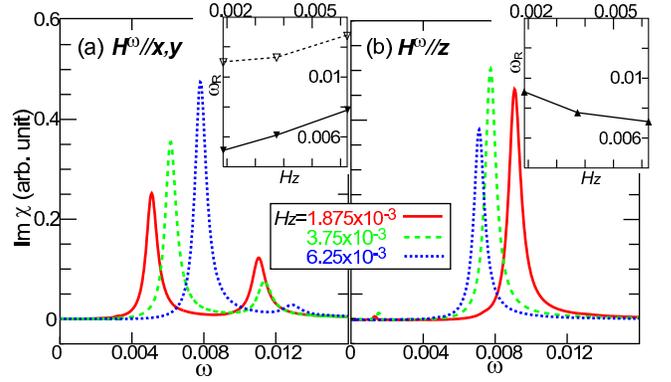}
\caption{(color online). Imaginary parts of (a) in-plane and (b) out-of-plane dynamical susceptibilities, Im$\chi(\omega)$, in the SkX phase for several values of $H_z$. Insets show resonant frequencies $\omega_{\rm R}$ as functions of $H_z$.}
\label{Fig02}
\end{figure}
We then study the microwave-absorption spectra due to spin-wave resonances in the SkX phase. We trace spin dynamics after applying a $\delta$-function pulse of magnetic field at $t$=0, which is given by $\bm H'(t)$=$\delta(t)\bm H^{\omega}$. The absorption spectrum or the imaginary part of the dynamical susceptibility, Im$\chi(\omega)$, is calculated from the Fourier transformation of magnetization $\bm m(t)$=$(1/N)\sum_i \bm S_i(t)$.

In Fig.~\ref{Fig02}(a), we show calculated spectra for several values of $H_z$ when $\bm H^{\omega}$ is parallel to the $xy$ plane. We find two resonance peaks in the spectra, and both of their frequencies increase as $H_z$ increases as shown in the inset of Fig.~\ref{Fig02}(a). Note that $\omega$=0.01 corresponds to $\sim$1 GHz for $J$=0.4 meV (=96.7 GHz). Thus these spin-wave resonances are located in the frequency range 500 MHz-1.2 GHz or in the microwave regime. On the other hand, the calculated spectra for $\bm H^{\omega}$ parallel to the $z$ axis are shown in Fig.~\ref{Fig02}(b), which have only one resonance peak. The resonant frequency $\omega_{\rm R}$ decreases as $H_z$ increases as shown in the inset. Again these resonances are located in the microwave frequency regime.

\begin{figure*}[tdp]
\includegraphics[scale=0.45]{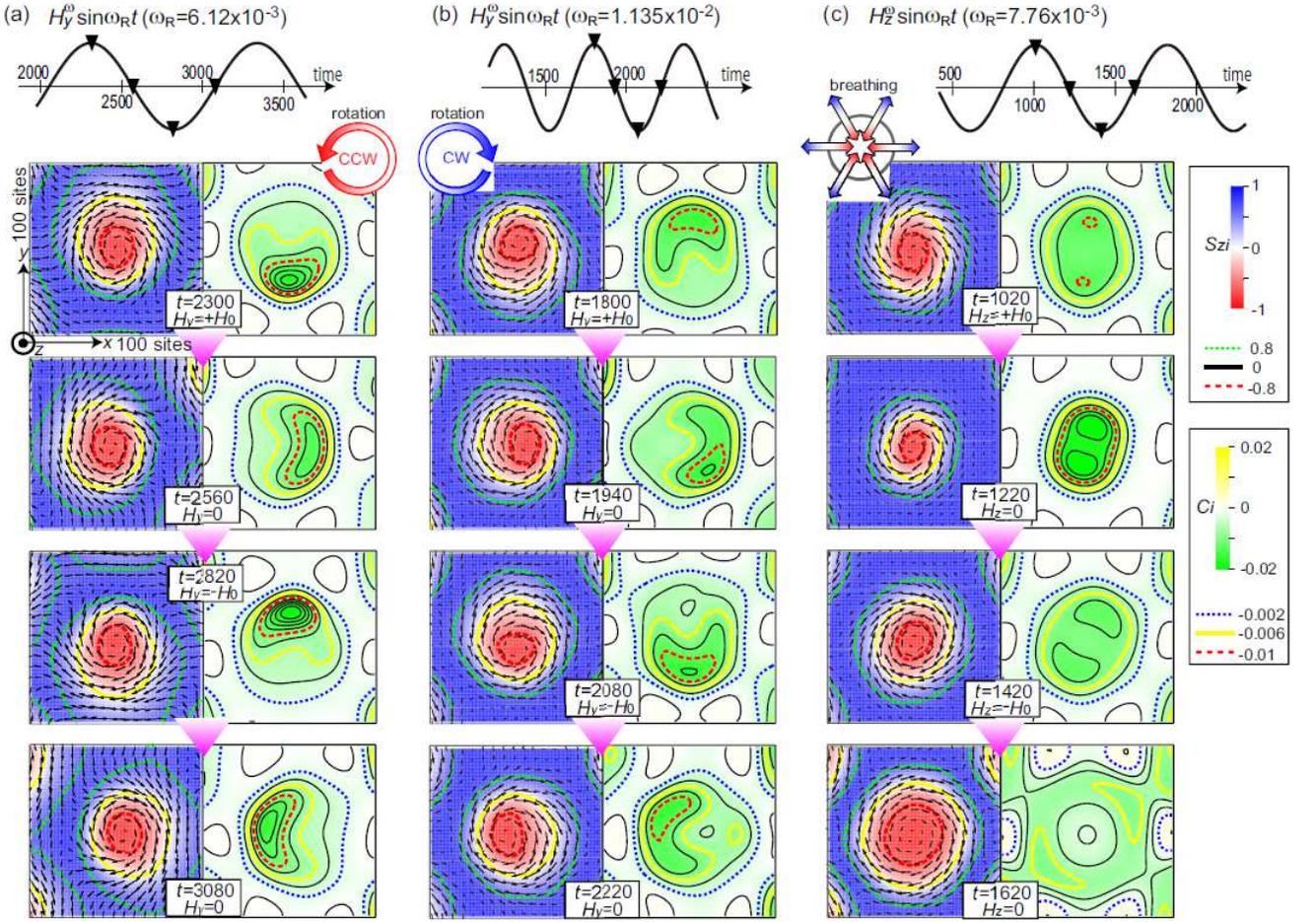}
\caption{(color). Spin dynamics of each collective mode in the SkX phase calculated at $H_z$=3.75$\times$10$^{-3}$. Spins at sites ($i_x$, $i_y$) are represented by arrows when mod($i_x$, 6)=mod($i_y$, 6)=0 with color maps of the $S_{zi}$ components in the left panels, while in the right panels, distributions of the local spin chiralities $C_i$ are displayed. Temporal waveforms of the applied a.c. magnetic fields, $H_y^{\omega}\sin\omega_{\rm R}t$ and $H_z^{\omega}\sin\omega_{\rm R}t$, are shown in the uppermost figures where inverted triangles indicate times at which we observe the spin configurations shown here. (a) [(b)] Lower-energy [Higher-energy] rotational mode with $\omega_{\rm R}$=6.12$\times$10$^{-3}$ ($\omega_{\rm R}$=1.135$\times$10$^{-2}$) activated by the {\it in-plane} a.c. magnetic field. Distributions of the $S_{zi}$ components and the spin chiralities $C_i$ circulate around the skyrmion core in a counterclockwise (clockwise) fashion. (c) Breathing mode with $\omega_{\rm R}$=7.76$\times$10$^{-3}$ activated by the {\it out-of-plane} a.c. magnetic field.}
\label{Fig03}
\end{figure*}
To identify each spin-wave mode, we trace the spin dynamics by applying a stationary oscillating magnetic field with resonant frequency $\omega_{\rm R}$. We first study the modes activated by the {\it in-plane} a.c. magnetic field by setting $\bm H'(t)$=(0, $H_y^{\omega}\sin\omega_{\rm R} t$, 0) with $H_y^{\omega}$=0.5$\times$10$^{-3}$. The frequency $\omega_{\rm R}$ is fixed at $\omega_{\rm R}$=6.12$\times$10$^{-3}$ for the lower-energy mode, while at $\omega_{\rm R}$=1.135$\times$10$^{-2}$ for the higher-energy mode. We find that for all of the modes, all the skyrmions show uniformly the same motion so that we focus on one skyrmion hereafter. In Figs.~\ref{Fig03}(a) and (b), we display calculated time evolutions of the spins. The spins at sites ($i_x$, $i_y$) are represented by arrows when mod($i_x$, 6)=mod($i_y$, 6)=0 together with distributions of the $S_{zi}$ components in the left panels, while those of the spin chiralities $C_i$ in the right panels. Interestingly the area of larger $S_{zi}$ or that of larger $|C_i|$ circulates around each skyrmion core even though the applied a.c. field $\bm H'(t)$ is linearly polarized in the $y$ direction. We find that directions of their rotations are opposite, i.e., counterclockwise (CCW) with respect to the magnetic field $\bm H$$\parallel$$\bm z$ for the lower-lying mode while clockwise (CW) for the higher-lying mode. These directions are independent of the sign of DM constant $D$ or winding direction of the spins. Instead they are determined by a sign of the applied field or by the spin orientation at the skyrmion core.

\begin{figure}
\includegraphics[scale=1.0]{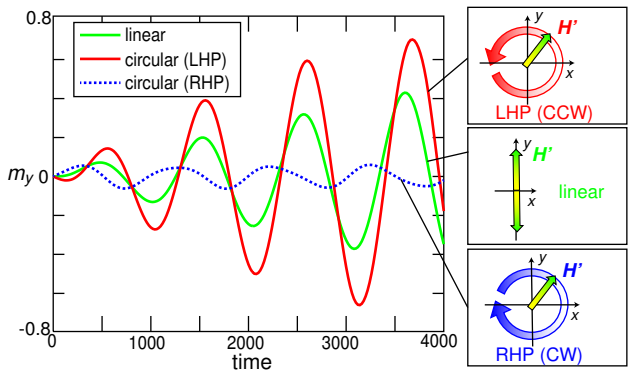}
\caption{(color online). Calculated time evolutions of magnetization ($\parallel$$\bm y$), $m_y(t)$=$(1/N)\sum_i S_{yi}(t)$, in the SkX phase at $H_z$=3.75$\times$10$^{-3}$ under linearly-polarized, left-handed circularly polarized (LHP), and right-handed circularly polarized (RHP) in-pane a.c. magnetic fields with resonant frequency $\omega_{\rm R}$=6.12$\times$10$^{-3}$ corresponding to the lower-lying mode.}
\label{Fig04}
\end{figure}
Because of these habits, the spin-wave excitations activated by the in-plane a.c. magnetic field strongly depend on the circular polarization of the irradiating microwave. In Fig.~\ref{Fig04}, we show calculated time evolutions of the magnetization parallel to the $y$ axis, $m_y(t)$=$(1/N)\sum_i S_{yi}(t)$, when we irradiate linearly-polarized, left-handed circularly polarized (LHP), and right-handed circularly polarized (RHP) in-plane microwaves with resonant frequency $\omega_{\rm R}$=6.12$\times$10$^{-3}$, which corresponds to the lower-lying mode at $H_z$=3.75$\times$10$^{-3}$. More concretely, we apply a time-dependent magnetic field $\bm H'(t)$=[$H'_x(t)$, $H'_y(t)$, 0] where $H'_x(t)$=$\alpha H_{xy}^{\omega}\cos\omega_{\rm R}t$ and $H'_y(t)$=$H_{xy}^{\omega}\sin\omega_{\rm R}t$ with $\alpha$=0 for the linearly-polarized microwave and $\alpha$=1 ($-$1) for the LHP (RHP) microwave. In the LHP (RHP) microwave, its magnetic-field component rotate in a CCW (CW) way. Here we fix $H_{xy}^{\omega}$=0.5$\times$10$^{-3}$. We find that irradiation of the LHP microwave significantly enhances the magnetization oscillation as compared to the linearly-polarized microwave, whereas the RHP microwave cannot activate collective spin oscillations. 

Next we discuss a spin-wave mode activated by the {\it out-of-plane} a.c. magnetic field. We again trace spin dynamics by applying $\bm H'(t)$=(0, 0, $H_z^{\omega}\sin\omega_{\rm R}t$) with $\omega_{\rm R}$=7.76$\times$10$^{-3}$ and $H_z^{\omega}$=0.5$\times$10$^{-3}$. We observe a breathing mode where the area of each skyrmion extends and shrinks dynamically as shown in Fig.~\ref{Fig03}(c). 

\begin{figure}
\includegraphics[scale=1.0]{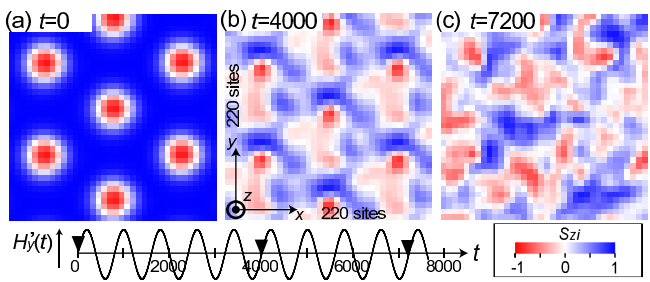}
\caption{(color online). Melting of the SkX within nano seconds under irradiating LHP microwave, which excites the rotational spin-wave modes intensively (see text). Color maps of the $S_{zi}$ components are displayed at (a) $t$=0, (b) $t$=4000, and (c) $t$=7200. Figures magnify a partial area with 220$\times$220 sites for clarity, while the calculations are done for 288$\times$288 sites with the periodic boundary condition. Temporal waveform of the microwave is also shown where times corresponding to figures (a), (b), and (c) are indicated by inverted triangles.}
\label{Fig05}
\end{figure}
We finally study effects of the intense spin-wave excitation. We apply in-plane LHP ($\alpha$=1) and RHP ($\alpha$=$-1$) microwaves of $H'_x(t)$=$\alpha H_{xy}^{\omega}\cos\omega t$ and $H'_y(t)$=$H_{xy}^{\omega}\sin\omega t$ to the SkX phase at $H_z$=6.3$\times$10$^{-3}$. The system is located on the phase boundary between the SkX and FM phases. Here we take $H_{xy}^{\omega}$=0.5$\times$10$^{-3}$, which corresponds to $\sim$1.7 mT when $J$=0.4 meV and $S$=1. The frequency $\omega$ is fixed at 7.4$\times$10$^{-3}$. This value is nearly equal to the resonant frequency $\omega_{\rm R}$=7.8$\times$10$^{-3}$ of the lower-energy mode, but slightly deviates from it in reality. Because the intense spin-wave excitations necessarily change the spin structure from its equilibrium configuration, and it results in red shifts of the resonant frequencies, we chose $\omega$ slightly smaller than $\omega_{\rm R}$ of the nearly equilibrium case in advance. In fact, the red shift can be seen in Fig.~\ref{Fig04}. The magnetization dynamics under the LHP microwave becomes slow as compared to that under the linearly-polarized microwave when the oscillation amplitude becomes larger. One can easily notice this fact from different maximum points between these two oscillations. Indeed the oscillation frequency in Fig.~\ref{Fig04} under the LHP microwave is $\omega$$\sim$6.1$\times$10$^{-3}$ for 0$<$$t$$<$2000, while $\omega$$\sim$5.7$\times$10$^{-3}$ for 3000$<$$t$$<$5000.

In Figs.~\ref{Fig05}(a) and (b), we show snap-shots of the spin configurations at several times under the irradiating LHP microwave. We observe melting of the SkX due to the intensively excited rotational spin-wave modes. The melting occurs within $t$$\sim$5000-6000. Here $t$=1000 corresponds to $\sim$10 nsec when $J$=0.4 meV. Thus the melting occurs within 50-60 nsec. We also find that the SkX melting is difficult to achieve either by the RHP microwave or even by the LHP microwave if its frequency is off-resonant. Note also that the spatial pattern in Fig.~\ref{Fig05}(c) loses a periodicity of the original SkX, suggestive of a chaotic aspect of the melting dynamics.

We finally compare the modes found in the SkX phase with those in the vortex-state nanodisks clarified in Refs.~\cite{Ivanov02,Zaspel05,Ivanov05,Zivieri08}. The twofold rotational modes and the breathing mode found in the SkX resemble, respectively, the twofold translational modes expressed by the Bessel functions with $m=\pm 1$ and the radial mode with $m=0$ in the vortex-state nanodisks. In Ref.~\cite{Ivanov05}, Ivanov and Zaspel theoretically showed that degeneracy of the translational modes with $m=\pm 1$ in the nanodisk is lifted under an applied magnetic field normal to the disk. We consider that a similar mechanism works in the SkX case for the doublet CW and CCW modes. There are also several differences. The modes in nanodisks are mainly governed by the long-range dipolar interaction, resulting in their salient aspect-ratio dependence. Note that their frequencies go to zero in the zero aspect-ratio limit. In contrast, the SkX and its dynamics considered here are governed by the nearest-neighbor spin interactions described in the Hamiltonian (1). The essential relevance of the DM interaction to the SkX is indicated by several experimental findings~\cite{YuXZ10N} such as its emergence only in chiral magnets, unique spin swirling directions of skyrmions, and considerably small size (10-100 nm) of skyrmions compared to dipolar-force-induced magnetic bubbles. Thus we expect negligible aspect-ratio dependence of the modes as well as weak influences of the dipolar interaction. Our study focuses on thin films whose thickness is much smaller than the skyrmion diameter because a greater stability of the SkX in thinner films has been confirmed~\cite{YiSD09,YuXZ10N}. In such a case, the system can be regarded as ferromagnetically stacked two-dimensional layers, which guarantees the validity of our results based on a two-dimensional model.

In summary, we have theoretically studied spin-wave excitations in the SkX phase of insulating ferromagnets with DM interaction. We have found a couple of rotational modes with $\sim$ 1 GHz frequencies for in-plane a.c. magnetic field. The rotations are in a CCW fashion for the lower-lying mode, while in a CW fashion for the higher-lying mode. These habits give rise to strong dependence of these spin-wave excitations on the circular polarization of the irradiating microwave. A breathing mode has been found for out-of-plane a.c. magnetic field. We have also observed the melting of the SkX under the irradiating LHP microwave. These findings will open a route to manipulation of skyrmion as a nano-scale spin texture using spin-wave resonances.

The author is deeply grateful to N. Nagaosa for fruitful discussion and insightful suggestions. The author also thanks Y. Tokura, M. Kawasaki, X. Z. Yu, and S. Seki for stimulating discussions. This work was supported by Grant-in-Aid (No. 22740214) and G-COE Program ``Physical Sciences Frontier" from MEXT Japan, Funding Program for World-Leading Innovative R$\&$D on Science and Technology (FIRST Program) on ``Quantum Science on Strong Correlation" from JSPS, and Strategic International Cooperative Program (Joint Research Type) from Japan Science and Technology Agency.



\begin{thebibliography}{999}
\bibitem{Malozemoff79}A. P. Malozemoff and J. C. Slonczewski, Magnetic domain walls in bubble materials, edited by R. Wolfe (Academic Press, New York, 1979).

\bibitem{BaderRMP06}S. D. Bader, Rev. Mod. Phys. {\bf 78}, 1 (2006).

\bibitem{Skyrme61}T. H. R. Skyrme, Proc. R. Soc. A {\bf 260}, 127 (1961).

\bibitem{Skyrme62}T. H. R. Skyrme, Nucl. Phys. {\bf 31}, 556 (1962).

\bibitem{Sondhi93}S. L. Sondhi, A. Karlhede, S. A. Kivelson, and E. H. Rezayi, Phys. Rev. B {\bf 47}, 16419 (1993).

\bibitem{Abolfath97}M. Abolfath, J. J. Palacios, H. A. Fertig, S. M. Girvin, and A. H. MacDonald, Phys. Rev. B {\bf 56}, 6795 (1997).

\bibitem{Heinze11}S. Heinze, K. von Bergmann, M. Menzel, J. Brede, A. Kubetzka, R. Wiesendanger, G. Bihlmayer, and S. Bl\"ugel, Nat. Phys. {\bf 7}, 713 (2011).

\bibitem{Raicevic11}I. Rai\v{c}evi\'c, Dragana Popovi\'c, C. Panagopoulos, L. Benfatto, M. B. Silva Neto, E. S. Choi, and T. Sasagawa, Phys. Rev. Lett. {\bf 106}, 227206 (2011).

\bibitem{Bogdanov89}A. N. Bogdanov and D. A. Yablonskii, Sov. Phys. JETP {\bf 68}, 101 (1989).

\bibitem{Bogdanov94}A. Bogdanov and A. Hubert, J. Mag. Mag. Mat. {\bf 138}, 255 (1994).

\bibitem{Muhlbauer09}S. M\"uhlbauer, B. Binz, F. Jonietz, C. Pfleiderer, A. Rosch, A. Neubauer, R. Georgii, P. B\"oni, Science {\bf 323}, 915 (2009).

\bibitem{Munzer10}W. M\"unzer, A. Neubauer, T. Adams, S. M\"uhlbauer, C. Franz, F. Jonietz, R. Georgii, P. B\"oni, B. Pedersen, M. Schmidt, A. Rosch, and C. Pfleiderer, Phys. Rev. B {\bf 81}, 041203(R) (2010).

\bibitem{Pfleiderer10}C. Pfleiderer, T. Adams, A. Bauer, W. Biberacher, B. Binz, F. Birkelbach, P. B\"oni, C. Franz, R. Georgii, M. Janoschek, F. Jonietz, T. Keller, R. Ritz, S. M\"uhlbauer, W. M\"unzer, A. Neubauer, B. Pedersen, and A. Rosch, J. Phys. Condens. Matter {\bf 22}, 164207 (2010).

\bibitem{YiSD09}S. D. Yi, S. Onoda, N. Nagaosa, and J. H. Han, Phys. Rev. B {\bf 80}, 054416 (2009).

\bibitem{YuXZ10N}X. Z. Yu, Y. Onose, N. Kanazawa, J. H. Park, J. H. Han, Y. Matsui, N. Nagaosa, and Y. Tokura, Nature (London) {\bf 465}, 901 (2010).

\bibitem{YuXZ10M}X. Z. Yu, N. Kanazawa, Y. Onose, K. Kimoto, W. Z. Zhang, S. Ishiwata, Y. Matsui, and Y. Tokura, Nat. Mat. (London) {\bf 10}, 106 (2010).

\bibitem{Jonietz10}F. Jonietz, S. M\"uhlbauer, C. Pfleiderer, A. Neubauer, W. M\"unzer, A. Bauer, T. Adams, R. Georgii, P. B\"oni, R. A. Duine, K. Everschor, M. Garst, and A. Rosch, Science {\bf 330}, 1648 (2010).

\bibitem{Everschor11}K. Everschor, M. Garst, R. A. Duine, and A. Rosch, Phys. Rev. B {\bf 84}, 064401 (2011).

\bibitem{SKHall}Current-driven motions of skyrmions have been recently studied theoretically: K. S. Kim and S. Onoda, arXiv:1012.0631; J. Zang, M. Mostovoy, J. H. Han, and N. Nagaosa, arXiv:1102.5384.



\bibitem{Binz08}B. Binz and A. Vishwanath, Physica B {\bf 403}, 1336 (2008).

\bibitem{LeeM07}M. Lee, Y. Onose, Y. Tokura, and N. P. Ong, Phys. Rev. B {\bf 75}, 172403 (2007).

\bibitem{Neubauer09}A. Neubauer, C. Pfleiderer, B. Binz, A. Rosch, R. Ritz, P. G. Niklowitz, and P. B\"oni, Phys. Rev. Lett. {\bf 102}, 186602 (2009).

\bibitem{Kanazawa11}N. Kanazawa, Y. Onose, T. Arima, D. Okuyama, K. Ohoyama, S. Wakimoto, K. Kakurai, S. Ishiwata, and Y. Tokura, Phys. Rev. Lett. {\bf 106}, 156603 (2011).

\bibitem{Pfleiderer10b}C. Pfleiderer and A. Rosch, Nature {\bf 465}, 880 (2010).
\bibitem{Bak80}P. Bak and M. H. Jensen, J. Phys. C {\bf 13}, L881 (1980).

\bibitem{Ivanov02}B. A. Ivanov and C. E. Zaspel, Appl. Phys. Lett. {\bf 81}, 1261 (2002).

\bibitem{Zaspel05}C. E. Zaspel, B. A. Ivanov, J. P. Park, and P. A. Crowell, Phys. Rev. B {\bf 72}, 024427 (2005).

\bibitem{Ivanov05}B. A. Ivanov and C. E. Zaspel, Phys. Rev. Lett. {\bf 94}, 027205 (2005).

\bibitem{Zivieri08}R. Zivieri and F. Nizzoli, Phys. Rev. B {\bf 78}, 064418 (2008).



\end{thebibliography}
\end{document}